\documentclass[a4paper,fleqn,usenatbib]{mnras}

\usepackage{graphicx}	
\usepackage{amsmath}	
\usepackage{amssymb}	
\usepackage{bm}

\title[Is there a fundamental $a_0$ in galaxies?]{Is there a fundamental acceleration scale in galaxies?}

\author[Z. Chang, Y. Zhou]{
Zhe Chang,$^{1,2}$
Yong Zhou$^{1,2}$\thanks{E-mail: zhouyong@ihep.ac.cn}
\\
$^{1}$Institute of High Energy Physics, Chinese Academy of Sciences, Beijing 100049, China\\
$^{2}$School of Physical Sciences, University of Chinese Academy of Sciences, Beijing 100049, China\\
}

\date{Accepted XXX. Received YYY; in original form ZZZ}

\pubyear{2019}

\begin{document}
\label{firstpage}
\pagerange{\pageref{firstpage}--\pageref{lastpage}}
\maketitle

\begin{abstract}
Milgrom's modified Newtonian dynamics (MOND) can explain well the mass discrepancy problem in galaxy without invoking dark matter. The MOND theory predicts a universal constant acceleration scale in galaxy, below which the Newtonian dynamics is no longer valid. Recently, Rodrigues et al. analyzed 193 high-quality disk galaxies by using Bayesian inference and they concluded that the probability of existence of a fundamental acceleration scale is essentially 0. In this paper, we use the same method to revisit the same question: Is there a fundamental acceleration scale in galaxies? We fit each galaxy rotation curve of 175 SPARC galaxies with Gaussian priors on galaxy parameters and a variable acceleration scale. After marginalizing over the nuisance parameters, we find that the marginalized posterior distributions of the acceleration scale become broad. The incompatibility between the global best fit and the marginalized posterior distributions of the acceleration scale is greatly reduced. However, there still exists evidence that rejects MOND as a fundamental theory. More accurate observations are needed to exclude or confirm the existence of a fundamental acceleration scale in galaxies.
\end{abstract}




\section{Introduction}
\label{Introduction}
In galactic scale, the mass discrepancy problem \citep{Rubin:1978kmz,Bosma:1981zz} has been found for many years. The approximately flatness of rotation curves in disk galaxies is faster than that expected from the luminous matter (stellar and gas). Therefore, it seems that there needs a significant amount of non-luminous matter, i.e. the dark matter in galaxy system. As the cornerstone of the the standard cosmological paradigm ($\Lambda$CDM), there are many strong indirect evidences that support the existence of dark matter. But up to now, no direct evidence of its existence has been found \citep{Akerib:2016vxi,Aprile:2017iyp,Cui:2017nnn}. A successful alternative to the dark matter hypothesis is the modified Newtonian dynamics \citep[MOND,][]{Milgrom:1983ca,Sanders:2002pf,Famaey:2011kh}, which attributes the mass discrepancy in galaxy system to a departure from the Newtonian dynamics at low accelerations. According to the MOND, the Newtonian dynamics is modified as the form,
\begin{eqnarray}\label{eq:MOND}
\bm{a}_{\textit{\tiny N}}=\mu\left(\frac{a}{a_0}\right)\bm{a}~~~or~~~\bm{a}=\nu\left(\frac{a_{\textit{\tiny N}}}{a_0}\right)\bm{a}_{\textit{\tiny N}},
\end{eqnarray}
where $\bm{a}_{\textit{\tiny N}}$ is the Newtonian acceleration, $\bm{a}$ is the kinematic acceleration, $\mu(x)$ or $\nu(y)$ is the interpolating function. The interpolating functions obey the relation $\mu(x)\nu(y)=1$ and have a characteristic acceleration scale $a_0$. The MOND theory predicts the acceleration scale with $a_0=\rm1.2\times10^{-13}~km~s^{-2}$, and there have two different asymptotic behaviors around this value. In the Newton region $a\gg a_0$, the Newtonian dynamics is recovered and $a=a_{\textit{\tiny N}}$. In the deep-MOND region $a\ll a_0$, the Newtonian dynamics is invalid and $a=\sqrt{a_0 a_{\textit{\tiny N}}}$. With an assumed constant acceleration scale, the MOND can explain well the flat rotation curves in outer part of disk galaxies. In particular, it naturally deduces the well known global scaling relation, the Tully-Fisher relation \citep{Tully:1977fu,McGaugh:2000sr,Lelli:2015wst}. Although many astronomical observations on disk galaxy show a universal constant acceleration scale, whether the acceleration scale is a fundamental constant or not, is still an open question. In fact, the acceleration scale is treated as a free parameter to fit the galaxy rotation curve, and different galaxy may somehow have different acceleration scale \citep{Begeman:1991iy,Swaters:2010qe,Chang:2013twa}.

Recently, \citet[hereafter R18]{Rodrigues:2018duc} analyzed 193 high-quality disk galaxies by using Bayesian inference and they concluded that the probability of existence of a fundamental acceleration scale is essentially 0. In particular, the MOND theory is ruled out as a fundamental theory for galaxies at more than $10\sigma$. However, their conclusions were built on flat priors on the galaxy parameters. The observational uncertainties of galaxy distance and inclination have been ignored \citep{2018NatAs...2..924M,Kroupa:2018kgv}, which have significant impact on the distribution of acceleration scale. It is interesting to note that there has been another study \citep[hereafter L18]{Li:2018tdo} that analyzed the same SPARC galaxy sample but found almost the opposite conclusion. They used an excessively strong Gaussian prior on acceleration scale which limits the variation of acceleration scale within a narrow region. The result based on a flat prior on acceleration scale is excluded by comparing the cumulative distributions functions (CDF) of reduced $\chi^2$. This method was doubted by \citet{Rodrigues:2018lvw}. In this paper, we use Bayesian inference to revisit the same question: Is there a fundamental acceleration scale in galaxies? It is argued that before Bayesian inference, one can't ignore the already existed information about the fitted parameter or take excessively strong prior. The flat priors on galaxy parameters are improper in this analysis. We use Gaussian  priors on galaxy parameters and a variable acceleration scale to fit each galaxy rotation curve of 175 SPARC galaxies. We obtain the posterior distribution of acceleration scale by marginalizing over the nuisance parameters. Then, we can compare the marginalized posterior distribution of acceleration scale for each galaxy with the global best fit of acceleration scale. If the global best fit is compatible with each galaxy, then the MOND theory could be still a fundamental theory and there exists a fundamental acceleration scale in galaxies.

The rest of this paper is organized as follows. In Section \ref{Method}, we make a brief introduction to the SPARC data set. Then we introduce the Bayesian analysis, especially the choice of Gaussian priors. In Section \ref{Result}, We analyse the MCMC result for galaxy NGC6195 and compare it with that from R18 and L18.  Then we compare the marginalized posterior distribution of acceleration scale for each galaxy with the global best fit of acceleration scale. Finally, conclusions and discussions are given in Section \ref{Conclusions}.

\section{Methodology}
\label{Method}
\subsection{SPARC data set}
\label{Data}
The \textit{Spitzer} Photometry and Accurate Rotation Curves (SPARC) data set \footnote{\url{http://astroweb.cwru.edu/SPARC/}} \citep{Lelli:2016zqa} is a sample of 175 disk galaxies with new surface photometry at $3.6~\mu$m and high-quality rotation curves from previous HI/H$\alpha$ studies. The surface photometry at $3.6~\mu$m provides the stellar mass via the mass-to-light ratio conversion factor. The gas mass is provided by the 21cm observations. Most of galaxies have disk component, only a few of galaxies have extra bulge component, both of them constitute the stellar component. In total, the galaxy baryonic mass profile comprises disk, bulge and gas component and we consider no dark matter. In SPARC data set, the mass profile is represented by velocity at a given radius, so the total baryonic velocity is
\begin{eqnarray}\label{eq:vbar}
V_{bar}^2=\Upsilon_{d}V^2_{disk}+\Upsilon_{b}V^2_{bulge}+V^2_{gas},
\end{eqnarray}
where $\Upsilon_{d}$ and $\Upsilon_{b}$ are the mass-to-light ratios for disk and bulge component, respectively. According to the MOND theory, the theoretical centripetal acceleration is tightly correlated with the total baryonic mass profile,
\begin{eqnarray}
\label{eq:gth1}
g_{th}&=&\frac{g_{bar}}{1-e^{-\sqrt{g_{bar}/a_0}}},
\qquad\qquad \text{RAR-inspired},\\
\label{eq:gth2}
g_{th}&=&g_{bar}\cdot\left(\frac{1}{2}+\sqrt{\frac{a_0}{g_{bar}}+\frac{1}{4}}\right),
\qquad \text{Simple},\\
\label{eq:gth3}
g_{th}&=&g_{bar}\cdot\sqrt{\frac{1}{2}+\sqrt{\frac{a^2_0}{g^2_{bar}}+\frac{1}{4}}},
\qquad \text{Standard},
\end{eqnarray}
where $g_{bar}=V^2_{bar}/R$ is the baryonic acceleration and the acceleration scale $a_0$ is treated as a free parameter to investigate its variation from galaxy to galaxy. The RAR-inspired, simple and standard interpolating functions are used here, respectively. The RAR-inspired one comes from the radical acceleration relation (RAR) \citep{McGaugh:2016leg} and the simple and standard interpolating functions are mostly used in early study on fitting galaxy rotation curves \citep{McGaugh:2008nc,Famaey:2011kh,Hees:2015bna}. Then we can deduce the theoretical rotation curve at every radius,
\begin{eqnarray}\label{eq:vth}
V_{th}=\sqrt{R\cdot g_{th}}.
\end{eqnarray}
Thus, we can compare the theoretical rotation curve with the observed rotation curve $V_{obs}$.

For individual galaxy, there are five factors that could affect the galaxy rotation curve. First, the stellar mass-to-light ratio $\Upsilon_\star$ could affect the total baryonic matter profile and then affect the theoretical rotation curve $V_{th}$. Second, uncertainties in galaxy distance affect the radius and the mass profile of baryonic components. If the galaxy distance $D$ is changed to $D'=D\delta_D$, where $\delta_D$ is a dimensionless distance factor, then the radius changes according to $R'=R\delta_D$ and the baryonic component velocity changes to $V'_k=V_k\sqrt{\delta_D}$, where $k$ denotes disk, bulge or gas. Third, uncertainties in galaxy inclination only affect the observed rotation curves and its uncertainties. If the galaxy inclination $i$ is changed to $i'=i\delta_i$, where $\delta_i$ is a dimensionless inclination factor, the observed rotation curves and its uncertainties change according to $V'_{obs}=V_{obs}\sin(i)/\sin(i')$ and $\delta V'_{obs}=\delta V_{obs}\sin(i)/\sin(i')$. Fourth, even the MOND theory predicts a universal constant acceleration scale, we treat the acceleration scale as a free parameter which affects the value of interpolating function $\nu(y)$ and then affects the theoretical rotation curve $V_{th}$. Fifth, the different interpolating function in equation (\ref{eq:gth1}-\ref{eq:gth3}) also affects theoretical rotation curve $V_{th}$. More details could be found in L18 or R18. An example of galaxy rotation curve for galaxy NGC6195 is shown in Fig. \ref{fig:NGC6195RC}.

\subsection{Bayesian analysis}
\label{Bayesian}

In order to fit the rotation curve for individual galaxy, we implement the Bayesian inference by using the affine-invariant Markov chain Monte Carlo (MCMC) ensemble sampler in $emcee$ \citep{2013PASP..125..306F}. The posterior probability of parameter space is $P(a_0,\Upsilon_{d},\Upsilon_{b},\delta_D,\delta_i|SPARC)=\mathcal{L}(a_0,\Upsilon_{d},\Upsilon_{b},\delta_D,\delta_i|SPARC)P(a_0,\Upsilon_{d},\Upsilon_{b},\delta_D,\delta_i)$, where the likelihood is derived from the $\chi^2$ function, $\mathcal{L}\sim e^{-\chi^2/2}$ and
\begin{eqnarray}\label{eq:chi2}
\chi^2=\sum_{k=1}^N \left(\frac{V_{th}(R_k;a_0,\Upsilon_{d},\Upsilon_{b},\delta_D)-V'_{obs,k}}{\delta V_{obs,k}^{'}}\right)^2,
\end{eqnarray}
where $N$ is the number of data point for individual galaxy, the observed rotation curve and its uncertaintiy at the radius $R_k$ has been changed to $V'_{obs,k}$ and $\delta V'_{obs,k}$ with a dimensionless factor $\delta_i$. The theoretical rotation curve $V_{th}$ at the radius $R_k$ is predicted by the MOND theory (with a given interpolating function) with given parameters $\{a_0,\Upsilon_{d},\Upsilon_{b},\delta_D\}$. The prior probability is the product of respective priors, $P(a_0,\Upsilon_{d},\Upsilon_{b},\delta_D,\delta_i)=P(a_0)P(\Upsilon_{d})P(\Upsilon_{b})P(\delta_D)P(\delta_i)$. Finally, we marginalize over the nuisance parameters set $\{\Upsilon_{d},\Upsilon_{b},\delta_D,\delta_i\}$, and the 1-d marginalized posterior of $a_0$ is as follow, $P(a_0|SPARC)=\int P(a_0,\Upsilon_{d},\Upsilon_{b},\delta_D,\delta_i|SPARC) d\Upsilon_{d}d\Upsilon_{b}d\delta_Dd\delta_i$.

\begin{figure}
\begin{center}
\includegraphics[width=\linewidth]{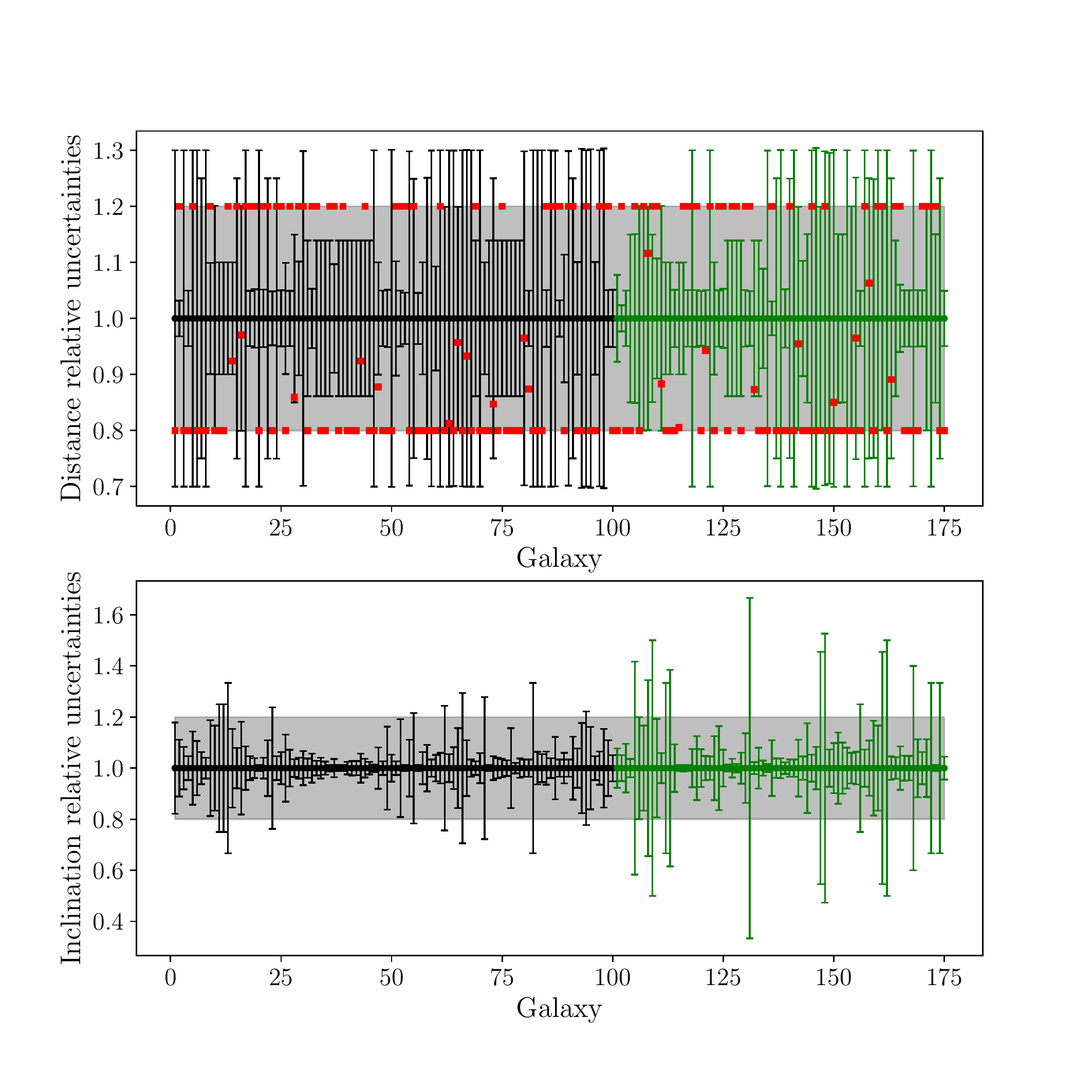}
\caption{The relative uncertainties of galaxy distance and inclination for 175 SPARC galaxies. The first 100 galaxies are selected by $5\sigma$ quality criteria in R18. The shaded regions represent relative uncertainty within $20\%$. The red dots mark the best fitting values of distance factor obtained from the minimization of $\chi^2$ in R18.}
\label{fig:uncertainty}
\end{center}
\end{figure}

The Bayesian inference needs priors that account all the information about the parameter space before analyzing the SPARC data set. In previous work, L18 used Gaussian priors on galaxy parameters while R18 used flat priors and they got almost the opposite conclusion. Fig. \ref{fig:uncertainty} illustrates the relative uncertainties of galaxy distance and inclination for 175 SPARC galaxies. In upper panel, the relative uncertainties of galaxy distance range from $5\%$ to $30\%$ according to different observational methods. The shaded region represents a tolerance of $20\%$ which is adopted as the range of fitting parameter in R18 and they found most of the best fitting values of distance factor locate at the boundary. This range regarded as a flat prior may be improper, the existed uncertainty is ignored and the truncation may affect the distribution of acceleration scale. In addition, their study ignored the uncertainty of galaxy inclination which also affect the distribution of acceleration scale. In L18, Gaussian priors were imposed with the uncertainties of galaxy distance and inclination. In this paper, we use the same Gaussian priors on the galaxy distance and inclination factor. The same case happens to the choice of prior on the stellar mass-to-light ratio. R18 took a tolerance of a factor of 2 on the stellar mass-to-light ratio while L18 imposed a Gaussian prior accounting for the stellar population synthesis (SPS) model \citep{2014PASA...31...36S}. We take the same Gaussian prior on the stellar mass-to-light ratio as L18. Before the Bayesian inference, there is no information about the acceleration scale for individual galaxy. One can't take excessively strong prior on it which limits the variation of acceleration scale within narrow region. Here we take a very weak Gaussian prior on the acceleration scale. In order to compare our result with R18 on the distribution of acceleration scale, we use the same notation that $a_0$ in the posterior probability stands for $\log_{10}a_0$. Specifically, the Gaussian prior on each parameter is represented as the normal distributions: $P(\delta_D)=\mathcal{N}(1,(\sigma_D/D)^2)$, $P(\delta_i)=\mathcal{N}(1,(\sigma_i/i)^2)$, $P(\Upsilon_{d})=\mathcal{N}(0.5,0.1^2)$, $P(\Upsilon_{b})=\mathcal{N}(0.7,0.15^2)$, $P(a_0)=\mathcal{N}(-12.921,2^2)$, where $\log_{10}(1.2\times10^{-13})\approx-12.921$. We also impose physical constraint on these parameters as follow: $\Upsilon_{d}>0,\Upsilon_{b}>0,\delta_D>0,90^\circ>i\delta_i>0^\circ,a_0>0$.

\section{Result}
\label{Result}

We use $emcee$ \citep{2013PASP..125..306F} to implement the Bayesian inference and initialize the MCMC chains with 100 random walkers. We run 500 steps in the burn-in phase and another 2000 steps in the production phase, which is enough for our purpose. We also check the acceptance fractions for all galaxies within the range (0.1, 0.7) as L18.

\begin{figure}
\begin{center}
\includegraphics[width=\linewidth]{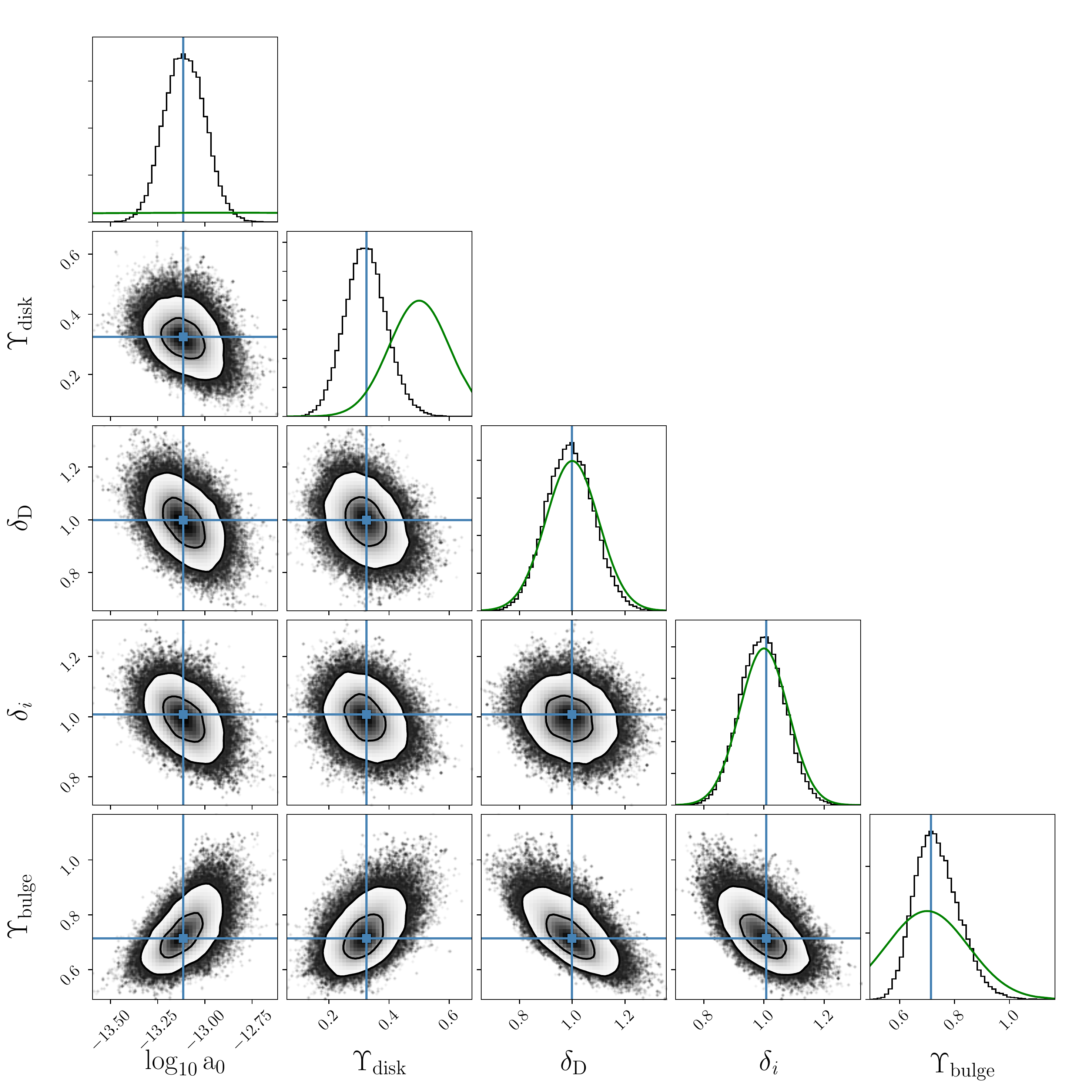}
\caption{The 1-dimensional and 2-dimensional marginalized posteriors on the parameter space $\{\log_{10} a_0,\Upsilon_{d},\Upsilon_{b},\delta_D,\delta_i\}$ for galaxy NGC6195. The horizontal and vertical solid lines mark the maximum of 1-dimensional marginalized posteriors. The contours from grey to light mark $1\sigma$ and $2\sigma$ credible regions of 2-dimensional marginalized posteriors. The green lines show the prior distributions on each parameter.}
\label{fig:NGC6195Tri}
\end{center}
\end{figure}

\begin{figure}
\begin{center}
\includegraphics[width=\linewidth]{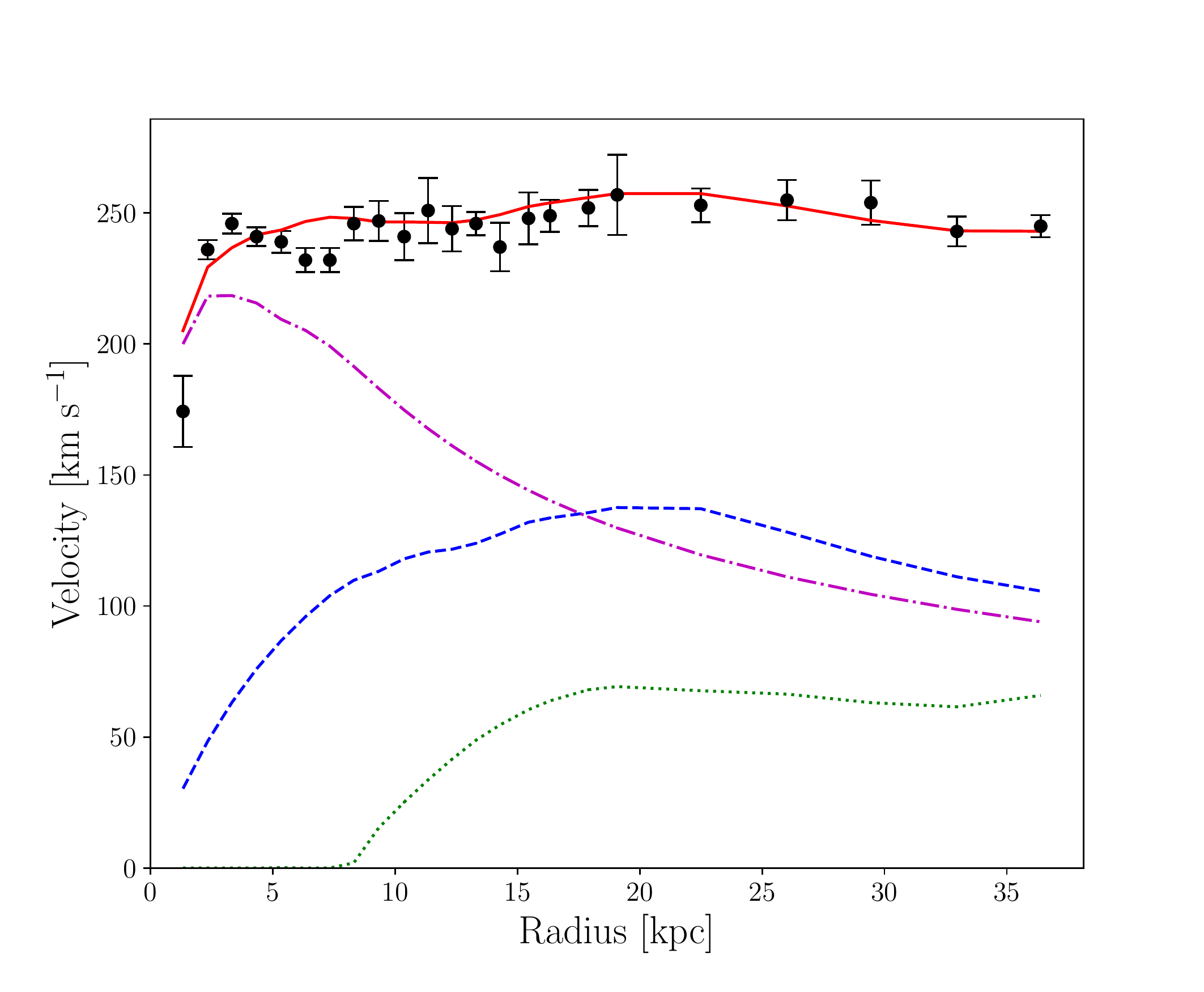}
\caption{Example of galaxy rotation curves for galaxy NGC6195. The radius and all circular velocities have been corrected according to the maximum of 1-dimensional marginalized posteriors from Fig. \ref{fig:NGC6195Tri}. The points with error bars are the observed rotation curves. The solid line represents the total baryonic velocity. Each baryonic component velocity is represented: dotted line for the gas, dashed line for the disk, and dash-dotted line for the bulge.}
\label{fig:NGC6195RC}
\end{center}
\end{figure}

Fig. \ref{fig:NGC6195Tri} shows the example of the Bayesian analysis with Gaussian priors for galaxy NGC6195 using the RAR-inspired interpolating function. The same Bayesian analysis of this galaxy is also represented in R18 and L18. In Fig. \ref{fig:NGC6195Tri}, we find that there is no strong correlation between the parameters when Gaussian priors are employed. It is the same with the result in Fig. A.1 of L18. But in Fig. 4 of R18 supplementary,  there showed strong correlations between the parameters when flat priors were employed. The flat priors have no constraint for whole parameter space and the posterior is only determined by the likelihood, which ignore the existed uncertainties and lead to the maximum of posterior on $\Upsilon_d$ locating at the boundary. It also happens to other parameter, especially for the distance factor $\delta_D$ in most galaxies. The Gaussian priors which account for the existed information are more reasonable to constrain the parameter to be physical and not far from the expected value. L18 showed the Bayesian analysis with Gaussian priors for the parameter space $\{\Upsilon_{d},\Upsilon_{b},\delta_D,\delta_i\}$ and fixed the acceleration scale at $a_0=\rm1.2\times10^{-13}~km~s^{-2}$. Fig. \ref{fig:NGC6195Tri} shows that a variable acceleration scale could reduce the variation of other parameter. The maximum of marginalized posterior is $\log_{10}a_0=-13.115,~\Upsilon_{d}=0.325,~\Upsilon_{b}=0.714,~\delta_D=0.999,~\delta_i=1.008$. In addition, it also shows that the Gaussian prior imposed on $a_0$ is almost flat respect to the marginalized posterior, there is extremely weak constraint on the distribution of $a_0$, which prevents $a_0$ too small. In Fig. \ref{fig:NGC6195RC}, we show the rotation curves for galaxy NGC6195 by using the maximum of marginalized posterior. A similar fit is obtained as well as L18.

\begin{figure*}
\centering
\begin{minipage}[b]{\linewidth}
\includegraphics[width=0.5\linewidth]{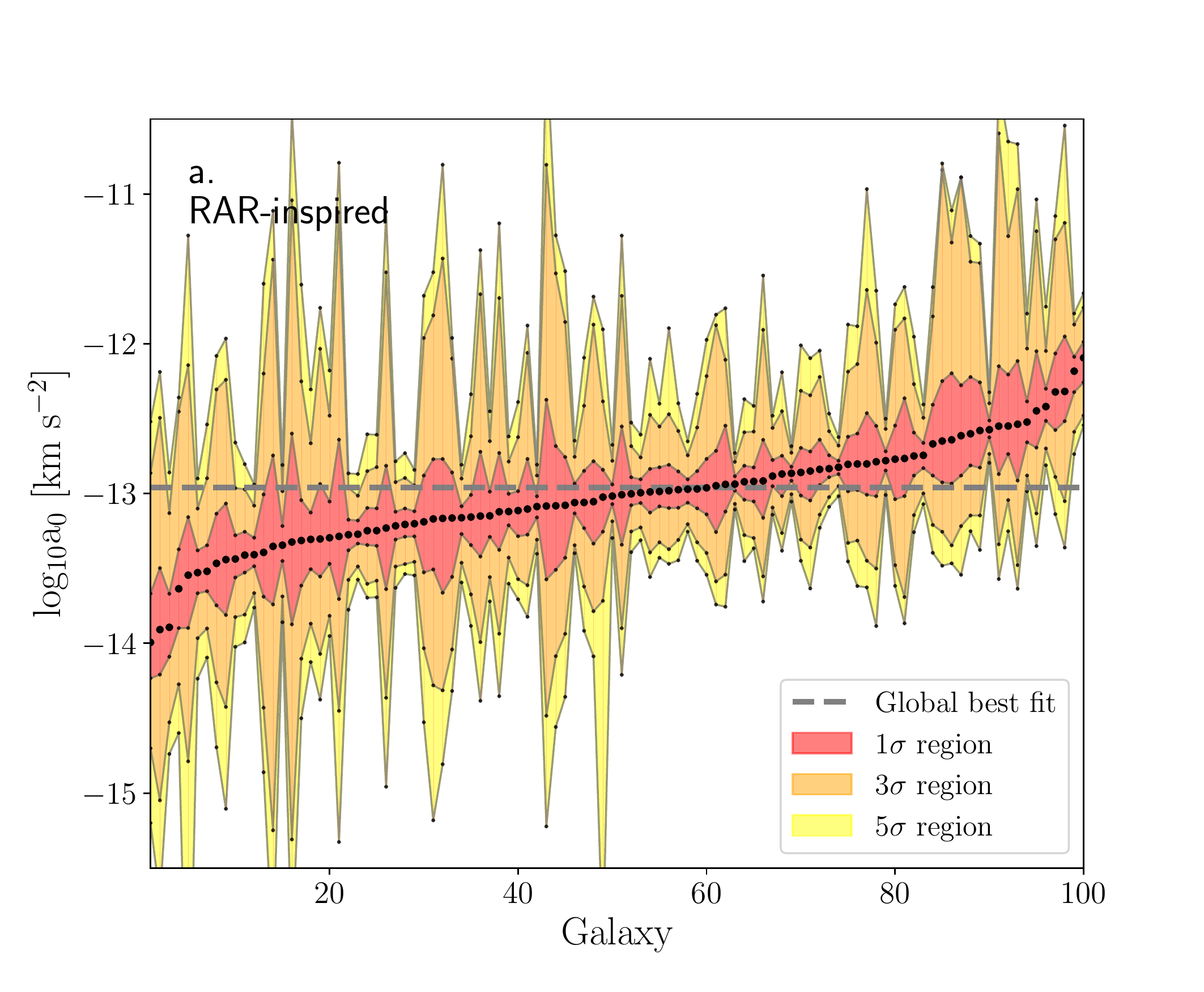}\vspace{1pt}
\includegraphics[width=0.5\linewidth]{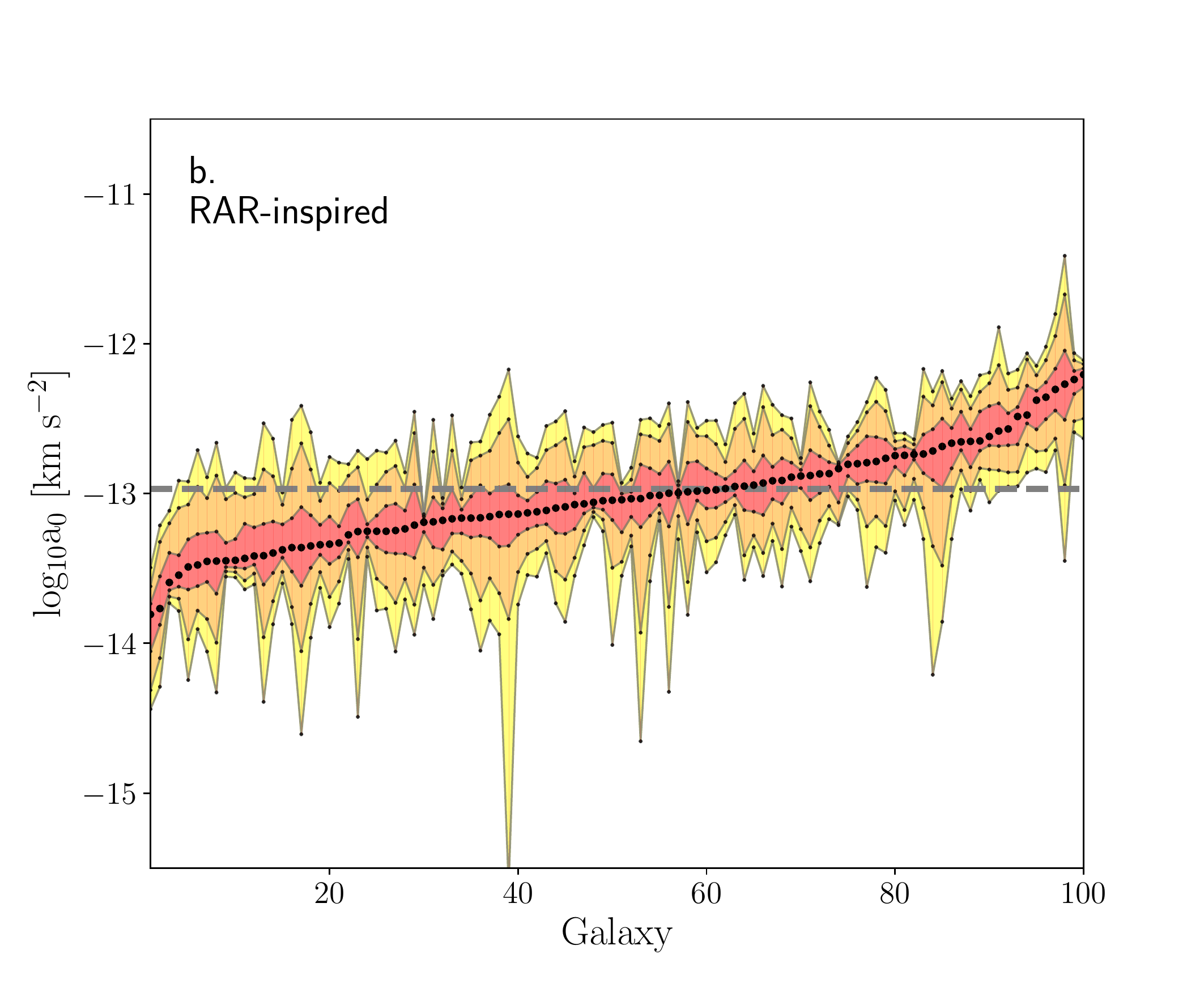}\
\includegraphics[width=0.5\linewidth]{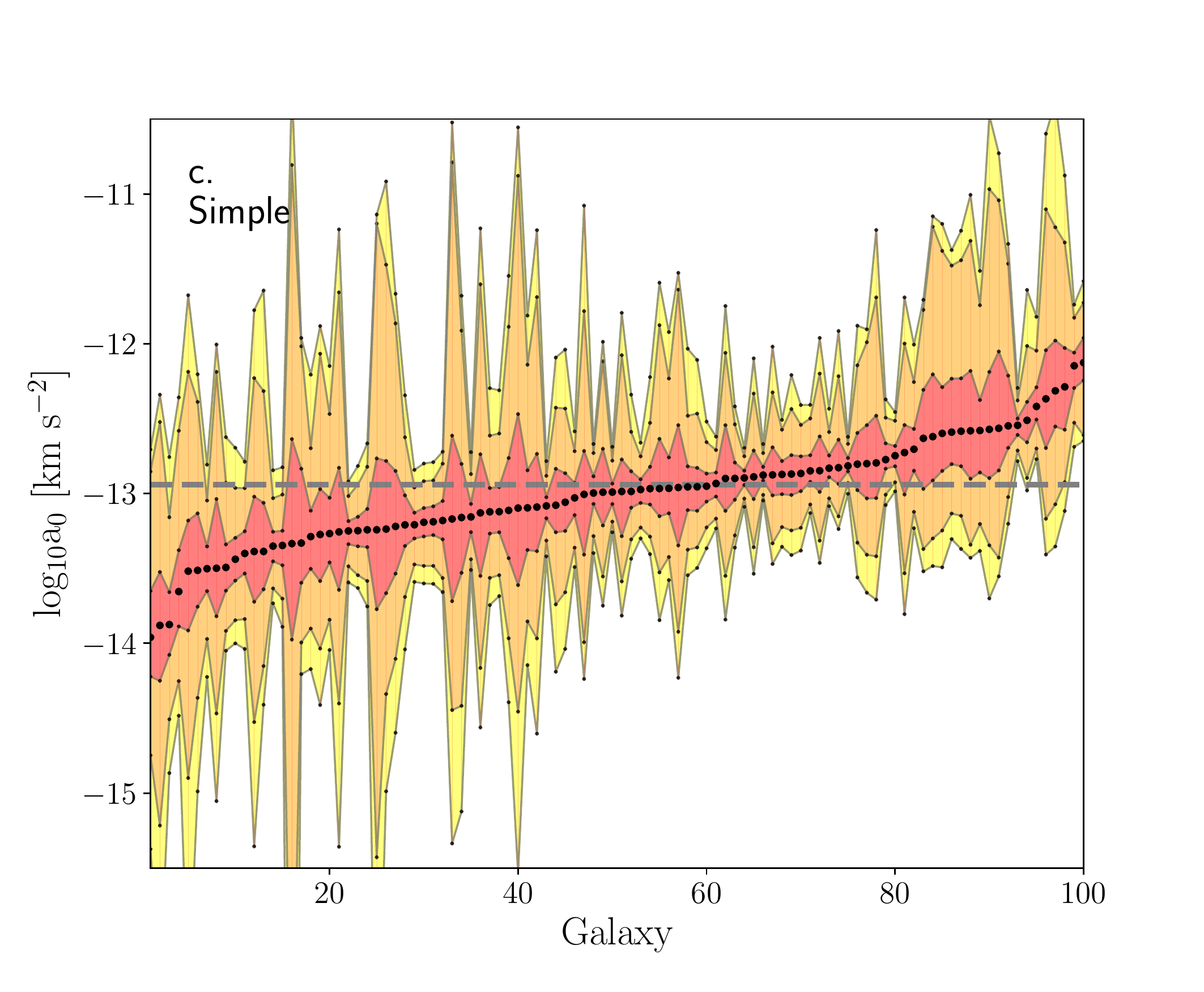}\vspace{1pt}
\includegraphics[width=0.5\linewidth]{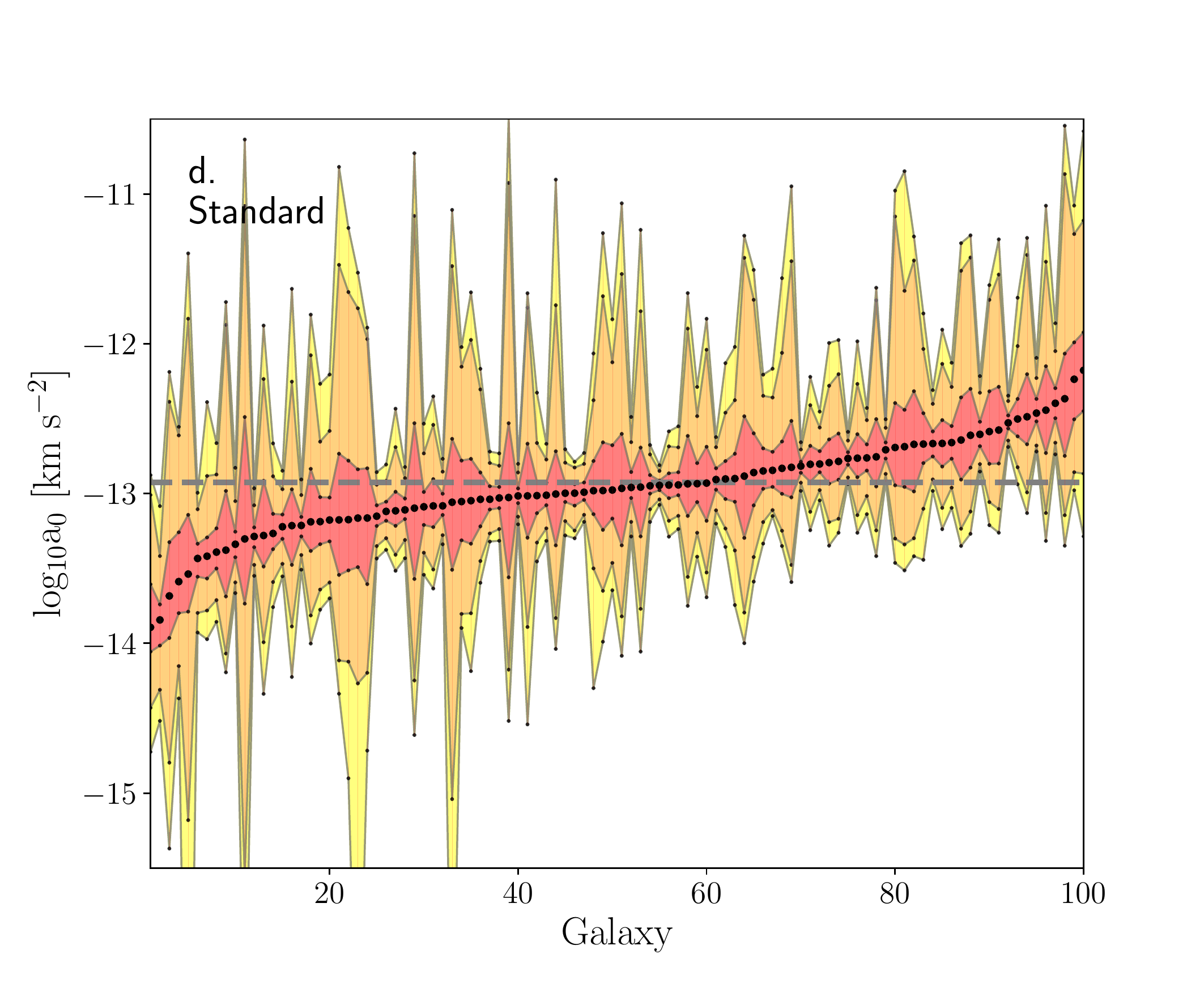}
\end{minipage}
\caption{The marginalized posterior distributions of $\log_{10} a_0$ for the SPARC galaxies. Only 100 galaxies passed the $5\sigma$ quality criteria are displayed. The black points mark the maximum of posterior and the red, orange and yellow shaded regions are the $1\sigma$, $3\sigma$ and $5\sigma$ confidence intervals, respectively. The grey dashed line indicates the global best fit of $\log_{10} a_0$. These galaxies are sorted by the maximum of marginalized posterior. Panels \textbf{a}, \textbf{c} and \textbf{d} respectively use the RAR-inspired, simple and standard interpolating functions with the same Gaussian priors. Panel $\textbf{b}$ shows the result of the RAR-inspired interpolating function with flat priors in R18.}
\label{fig:main}
\end{figure*}

\begin{figure}
\begin{center}
\includegraphics[width=\linewidth]{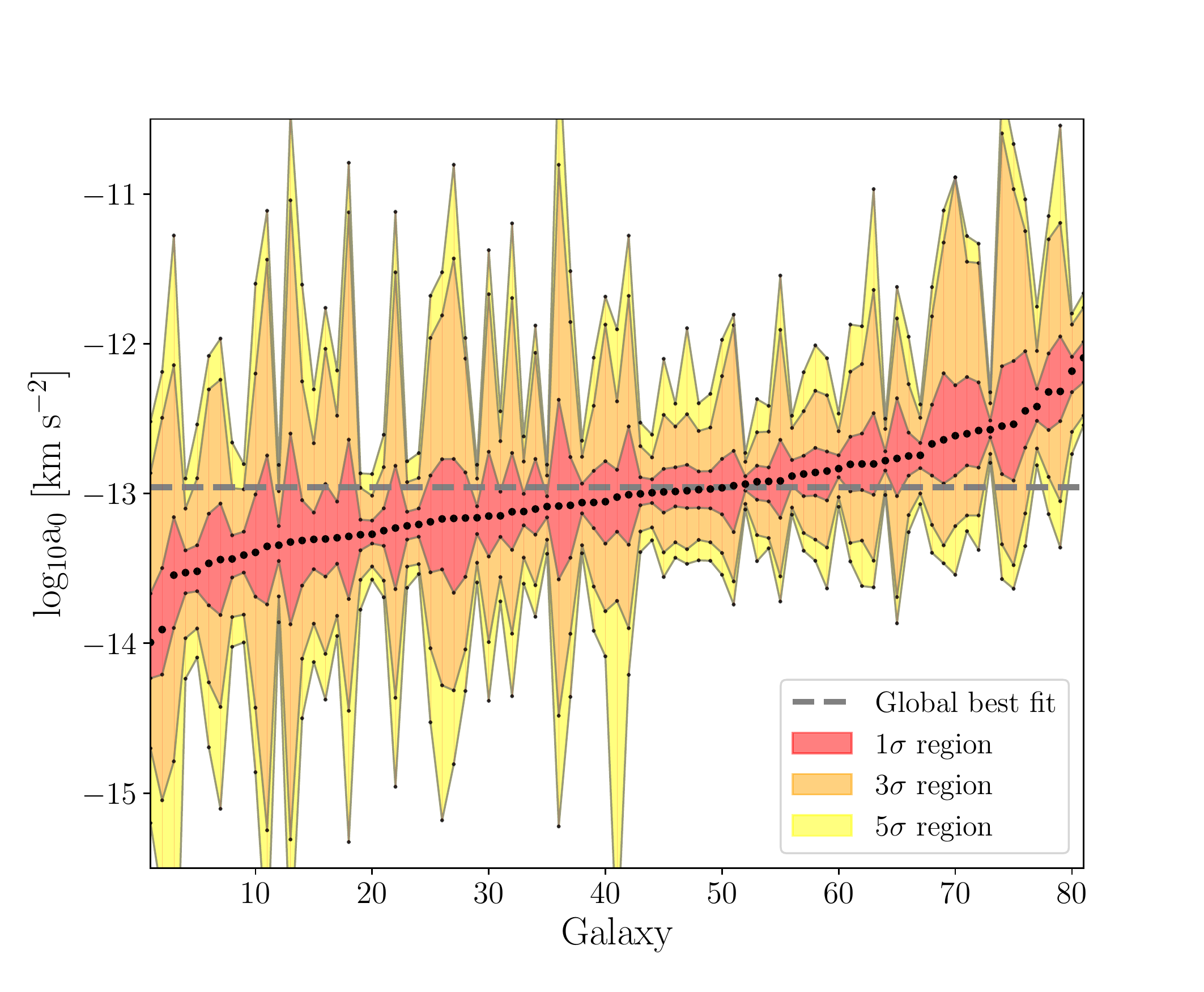}
\caption{The marginalized posterior distributions of $\log_{10} a_0$ for the RAR-inspired interpolating function. Only 81 galaxies passed the $3\sigma$ quality criteria are displayed.}
\label{fig:RARFivePlot81}
\end{center}
\end{figure}

The same Bayesian analysis with Gaussian priors for all 175 SPARC galaxies are also implemented. In order to obtain the marginalized posterior distribution of $a_0$ for individual galaxy, we marginalize over the nuisance parameters $\{\Upsilon_{d},\Upsilon_{b},\delta_D,\delta_i\}$. In R18, four quality cuts had been performed to exclude the galaxy with poor quality or poor fit. Only 100 (81) SPARC galaxies passed the $5\sigma$ ($3\sigma$) quality criteria. These quality criteria together with the maximum of marginalized posterior of $a_0$ and its $1\sigma$, $3\sigma$ and $5\sigma$ confidence intervals for the RAR-inspired interpolating function are summarized in Table \ref{tab:a0}. Fig. \ref{fig:main} shows the marginalized posterior distributions of $a_0$ for those galaxies passed $5\sigma$ quality criteria. When Gaussian priors are used and the galaxy inclination is considered for the Bayesian analysis, we find that the marginalized posterior distributions of $a_0$ become broader than that in R18, especially for the high or low acceleration range. This feature is held for the simple and standard interpolating functions. We also perform the global best fit and estimate the confidence level in rejecting MOND as employed in R18. As shown in Fig. \ref{fig:main}, the percentages of galaxies whose $5\sigma$ confidence intervals are incompatible with the global best fit is greatly reduced when Gaussian priors are employed. The statistical result is described in Table \ref{tab:bayesian}. There are only 4 galaxies comparing to 15 galaxies in R18, among the selected 100 galaxies, are incompatible with the global best fit by more than $5\sigma$. Most of these galaxies have large acceleration scale. At low acceleration range, those galaxies among 15 incompatible galaxies in R18 are compatible with the global best fit when Gaussian priors are employed, even though three galaxies are still incompatible for the standard interpolating function, but the discrepancy is greatly reduced. Using any of the three interpolating functions, the confidence level in rejecting MOND with Gaussian priors is nearly $25\sigma$, about half of $50\sigma$ in R18. Fig. \ref{fig:RARFivePlot81} shows the marginalized posterior distributions of $a_0$ for those galaxies passed $3\sigma$ quality criteria. There are only 4 galaxies, among the selected 81 galaxies, are incompatible with the global best fit by more than $5\sigma$. The confidence level in rejecting MOND with Gaussian priors drops to $22\sigma$, about half of $45\sigma$ in R18.

\begin{table*}
\caption{Global best fitting values of $\log_{10} a_0$ for each interpolating function with Gaussian priors and the percentages of galaxies whose $1\sigma$, $3\sigma$ and $5\sigma$ confidence intervals on $\log_{10} a_0$ are incompatible with the global best fit. The results of R18 with flat priors are listed as reference.}
\label{tab:bayesian}
\begin{center}
\begin{tabular}{lccccccccc}
\hline
\hline
  Interpolating & \multicolumn{4}{c}{Gaussian priors}                                                  && \multicolumn{4}{c}{Flat priors} \\ \cline{2-5} \cline{7-10}
  function      & $\log_{10} a_0 $ & $1\sigma$ ($32\%$)&$3\sigma$ ($0.27\%$)&$5\sigma$ ($0.000057\%$) && $\log_{10} a_0 $ & $1\sigma$ ($32\%$)&$3\sigma$ ($0.27\%$)&$5\sigma$ ($0.000057\%$)\\
  \hline
  Standard      &-12.925 & 56\% & 18\%& 8\% &&-12.899 & 70\%& 33\%& 20\% \\
  Simple        &-12.941 & 55\% & 17\%& 4\% &&-12.954 & 70\%& 34\%& 16\% \\
  RAR-inspired  &-12.959 & 55\%& 15\% & 4\% &&-12.970 & 71\%& 31\%& 15\% \\
\hline
\hline
\end{tabular}
\end{center}
\end{table*}

\begin{figure}
\centering
\includegraphics[width=\linewidth]{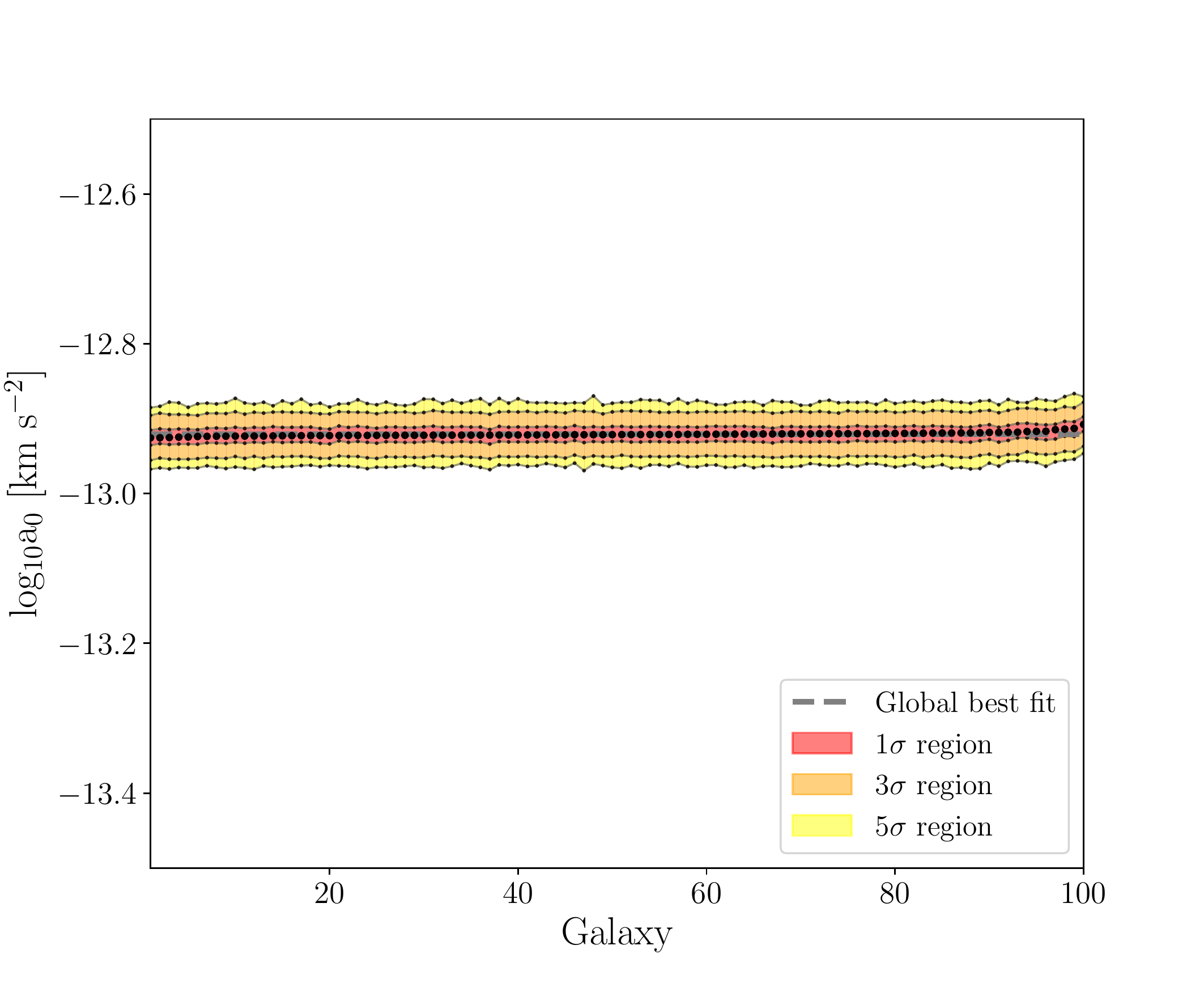}
\caption{The marginalized posterior distributions of $\log_{10} a_0$ for the RAR-inspired interpolating function. Here we impose an excessively strong Gaussian prior $P(a_0)=\mathcal{N}(-12.921,0.01^2)$ as same as L18.}
\label{fig:RARFivePlotnarrow}
\end{figure}

In L18, they also investigated the question whether the value of acceleration scale is truly a constant. They made Bayesian analysis with both a flat prior and a Gaussian prior on the acceleration scale. The Gaussian prior they used is $a_0=\rm(1.2\pm0.02)\times10^{-13}~km~s^{-2}$, which comes from the overall fitting of RAR for all SPARC galaxies \citep{McGaugh:2016leg}. For individual galaxy, it may be unreasonable to take the overall fitting result as their prior. Fig. \ref{fig:RARFivePlotnarrow} shows the marginalized posterior distributions of $a_0$ when an excessively strong Gaussian prior on $a_0$ is employed. Comparing to panel $\textbf{a}$ in Fig. \ref{fig:main}, we find that the tight distribution of $a_0$ originates from the excessively strong Gaussian prior. By comparing the cumulative distributions functions (CDF) of reduced $\chi^2$ (only consider the likelihood), L18 excluded the result based on the flat prior. However, the best fitting value of each parameter is obtained by maximizing the posterior, not the likelihood. As discussed in \citet{Rodrigues:2018lvw}, the CDF of the free $a_0$ has an advantage over the fixed $a_0$ when the prior is included in reduced $\chi^2$. Therefore, it seems that there is still room for the variation of acceleration scale for individual galaxy.

\section{Conclusions and discussions}
\label{Conclusions}

In this paper, we used Bayesian inference to revisit the question: Is there a fundamental acceleration scale in galaxies?  In previous work, L18 used Gaussian priors on galaxy parameters while R18 used flat priors and they got almost the opposite conclusion. We showed that the flat priors in R18 may affect the distribution of acceleration scale $a_0$. We used the same Gaussian priors on galaxy parameters as L18 and treated the acceleration scale as a free parameter to fit each galaxy rotation curve in SPARC data set. A similar fit is obtained for each galaxy rotation curve as well as L18. When Gaussian priors on galaxy parameters are used, we found that the marginalized posterior distributions of $a_0$ become broader than that in R18, especially for the high or low acceleration range. The incompatibility between the global best fit and the marginalized posterior distributions of $a_0$ is greatly reduced. These results are held for all interpolation functions we used. For the RAR-inspired interpolation function, there are only 4 galaxies comparing to 15 galaxies in R18, among the selected 100 galaxies, are incompatible with the global best fit by more than $5\sigma$. The confidence level in rejecting MOND with Gaussian priors is nearly $25\sigma$, about half of $50\sigma$ in R18. Besides, we found that the tight distributions of $a_0$ in L18 originates from the excessively strong Gaussian prior on $a_0$. When the prior is included in reduced $\chi^2$,  \citet{Rodrigues:2018lvw} showed that the CDF of the free $a_0$ has an advantage over the fixed $a_0$. Therefore, there is still room for the variation of $a_0$ for individual galaxy.

Although the Gaussian priors on galaxy parameters are employed and the incompatibility between the global best fit and the marginalized posterior distributions of $a_0$ is greatly reduced. There still exist evidence (about half of that in R18) that rejects MOND as a fundamental theory. The existence of $a_0$ could be an emergent nature as stated in R18. In our previous work, we find the spatial distribution of $a_0$ could be anisotropic \citep{Zhou:2017lwy,Chang:2018vxs}, a dipole correction for $a_0$ could reduce the incompatibility in this paper to some extent. Furthermore, as clarified by R18, the SPARC data set are not raw observational data, they depend on galaxy surface brightness decomposition, interpretations of the inclinations and distances, assumptions on axial symmetry and choices for data binning. All of these factors could impact the SPARC data set and then impact the distribution of $a_0$ to some extent. More accurate observations for galaxy rotation curve and baryonic mass profile are needed to exclude or confirm the existence of a fundamental acceleration scale in galaxies.

\section*{Acknowledgements}

We thank the anonymous referee for valuable suggestions and comments. We are grateful to Xin Li, Hai-Nan Lin and Zhi-Chao Zhao for useful discussions. We are also thankful for the open access of the SPARC data set. This work is supported by the National Natural Science Fund of China under grant Nos. 11675182 and 11690022.



\providecommand{\noopsort}[1]{}\providecommand{\singleletter}[1]{#1}%




\appendix

\section{Additional material}

\begin{table*}
\caption{The maximum of marginalized posterior of $\log_{10} a_0$ and its $1\sigma$, $3\sigma$ and $5\sigma$ confidence intervals for the RAR-inspired interpolating function. All 175 SPARC galaxies are represented. The column $P_5$ ($P_3$) labelling with 1 indicates the galaxy passed the $5\sigma$ ($3\sigma$) quality criteria in R18.}
\label{tab:a0}
\begin{center}
\begin{tabular}{l|ccccccccc}
\hline
\hline
  Galaxy          &$P_5$&$P_3$&$\log_{10} a_0 $ & $1\sigma_-$ & $1\sigma_+$ & $3\sigma_-$ & $3\sigma_+$ & $5\sigma_-$ & $5\sigma_+$\\
  \hline
  CamB            &  0  &  0  &  -14.667  &   -0.272  &    0.335  &   -3.476  &    0.940  &   -6.344  &    1.120 \\
  D512-2          &  1  &  1  &  -13.230  &   -0.408  &    0.416  &   -1.133  &    1.707  &   -1.727  &    2.111 \\
  D564-8          &  1  &  1  &  -13.519  &   -0.133  &    0.174  &   -0.381  &    0.621  &   -0.576  &    0.980 \\
  D631-7          &  0  &  0  &  -13.083  &   -0.084  &    0.066  &   -0.215  &    0.237  &   -0.289  &    0.353 \\
  DDO064          &  1  &  1  &  -13.122  &   -0.255  &    0.393  &   -0.814  &    1.427  &   -1.231  &    1.926 \\
  DDO154          &  1  &  0  &  -13.017  &   -0.054  &    0.077  &   -0.168  &    0.235  &   -0.280  &    0.342 \\
  DDO161          &  1  &  1  &  -12.766  &   -0.251  &    0.402  &   -0.925  &    0.935  &   -1.100  &    1.145 \\
  DDO168          &  0  &  0  &  -13.001  &   -0.156  &    0.157  &   -0.400  &    0.551  &   -0.531  &    0.831 \\
  DDO170          &  1  &  0  &  -13.636  &   -0.262  &    0.263  &   -0.638  &    1.183  &   -0.963  &    1.277 \\
  ESO079-G014     &  1  &  0  &  -12.771  &   -0.267  &    0.224  &   -0.708  &    0.864  &   -0.846  &    1.034 \\
  ESO116-G012     &  1  &  1  &  -12.601  &   -0.211  &    0.379  &   -0.546  &    1.148  &   -0.650  &    1.320 \\
  ESO444-G084     &  1  &  1  &  -12.448  &   -0.245  &    0.398  &   -0.685  &    1.200  &   -0.903  &    1.412 \\
  ESO563-G021     &  0  &  0  &  -12.883  &   -0.087  &    0.087  &   -0.243  &    0.274  &   -0.316  &    0.419 \\
  F561-1          &  0  &  0  &  -14.495  &   -1.249  &    1.269  &   -4.967  &    4.010  &   -7.230  &    5.454 \\
  F563-1          &  0  &  0  &  -12.717  &   -0.443  &    0.335  &   -1.062  &    1.377  &   -1.506  &    1.840 \\
  F563-V1         &  0  &  0  &  -14.719  &   -0.832  &    0.745  &   -4.579  &    1.619  &   -7.600  &    2.270 \\
  F563-V2         &  0  &  0  &  -12.541  &   -0.558  &    0.583  &   -1.441  &    2.095  &   -2.253  &    2.540 \\
  F565-V2         &  1  &  1  &  -12.802  &   -0.206  &    0.339  &   -0.646  &    1.162  &   -0.824  &    1.835 \\
  F567-2          &  0  &  0  &  -13.657  &   -0.961  &    1.122  &   -4.901  &    3.500  &   -8.039  &    4.329 \\
  F568-1          &  0  &  0  &  -12.524  &   -0.276  &    0.419  &   -0.851  &    1.393  &   -1.148  &    1.955 \\
  F568-3          &  1  &  0  &  -13.082  &   -0.427  &    0.399  &   -1.004  &    1.551  &   -1.477  &    1.805 \\
  F568-V1         &  1  &  1  &  -12.550  &   -0.320  &    0.400  &   -0.789  &    1.955  &   -1.021  &    2.272 \\
  F571-8          &  0  &  0  &  -11.080  &   -0.179  &    0.162  &   -0.440  &    0.678  &   -0.546  &    0.854 \\
  F571-V1         &  1  &  1  &  -13.286  &   -0.418  &    0.646  &   -1.165  &    2.163  &   -2.041  &    2.494 \\
  F574-1          &  1  &  1  &  -12.962  &   -0.177  &    0.193  &   -0.435  &    0.746  &   -0.580  &    0.988 \\
  F574-2          &  0  &  0  &  -14.613  &   -1.255  &    0.976  &   -4.806  &    3.459  &   -6.615  &    4.597 \\
  F579-V1         &  0  &  0  &  -15.237  &   -1.120  &    0.892  &   -4.564  &    2.753  &   -6.987  &    3.762 \\
  F583-1          &  1  &  1  &  -12.916  &   -0.246  &    0.275  &   -0.637  &    1.009  &   -0.805  &    1.371 \\
  F583-4          &  1  &  1  &  -13.079  &   -0.350  &    0.322  &   -0.858  &    1.224  &   -1.279  &    1.563 \\
  IC2574          &  0  &  0  &  -13.065  &   -0.062  &    0.096  &   -0.169  &    0.330  &   -0.210  &    0.468 \\
  IC4202          &  0  &  0  &  -12.369  &   -0.262  &    0.380  &   -0.641  &    1.081  &   -0.866  &    1.148 \\
  KK98-251        &  1  &  1  &  -13.908  &   -0.300  &    0.411  &   -1.140  &    1.414  &   -1.791  &    1.720 \\
  NGC0024         &  1  &  1  &  -12.745  &   -0.086  &    0.082  &   -0.255  &    0.250  &   -0.326  &    0.339 \\
  NGC0055         &  1  &  1  &  -13.087  &   -0.072  &    0.068  &   -0.221  &    0.207  &   -0.315  &    0.279 \\
  NGC0100         &  1  &  1  &  -12.614  &   -0.266  &    0.337  &   -0.604  &    1.725  &   -0.929  &    1.725 \\
  NGC0247         &  1  &  0  &  -13.202  &   -0.085  &    0.083  &   -0.254  &    0.255  &   -0.345  &    0.359 \\
  NGC0289         &  1  &  1  &  -13.528  &   -0.138  &    0.148  &   -0.438  &    0.427  &   -0.708  &    0.628 \\
  NGC0300         &  1  &  1  &  -12.642  &   -0.292  &    0.444  &   -0.705  &    1.317  &   -0.824  &    1.531 \\
  NGC0801         &  0  &  0  &  -13.553  &   -0.058  &    0.056  &   -0.172  &    0.178  &   -0.251  &    0.261 \\
  NGC0891         &  0  &  0  &  -12.610  &   -0.048  &    0.055  &   -0.147  &    0.162  &   -0.183  &    0.264 \\
  NGC1003         &  0  &  0  &  -11.919  &   -0.327  &    0.354  &   -0.986  &    0.921  &   -1.107  &    0.981 \\
  NGC1090         &  1  &  0  &  -13.248  &   -0.096  &    0.152  &   -0.355  &    0.398  &   -0.447  &    0.644 \\
  NGC1705         &  0  &  0  &  -12.405  &   -0.095  &    0.107  &   -0.269  &    0.432  &   -0.378  &    0.702 \\
  NGC2366         &  0  &  0  &  -13.242  &   -0.089  &    0.087  &   -0.239  &    0.299  &   -0.328  &    0.558 \\
  NGC2403         &  0  &  0  &  -12.026  &   -0.093  &    0.076  &   -0.415  &    0.258  &   -0.468  &    0.299 \\
  NGC2683         &  1  &  1  &  -13.216  &   -0.092  &    0.094  &   -0.272  &    0.292  &   -0.414  &    0.430 \\
  NGC2841         &  1  &  1  &  -12.937  &   -0.043  &    0.052  &   -0.132  &    0.149  &   -0.170  &    0.208 \\
  NGC2903         &  0  &  0  &  -12.395  &   -0.080  &    0.088  &   -0.316  &    0.245  &   -0.377  &    0.285 \\
  NGC2915         &  1  &  1  &  -12.420  &   -0.095  &    0.119  &   -0.280  &    0.371  &   -0.392  &    0.667 \\
  NGC2955         &  0  &  0  &  -12.918  &   -0.117  &    0.131  &   -0.371  &    0.378  &   -0.485  &    0.593 \\
  NGC2976         &  0  &  0  &  -14.905  &   -0.975  &    1.164  &   -4.683  &    2.431  &   -7.620  &    2.818 \\
  NGC2998         &  1  &  0  &  -13.408  &   -0.077  &    0.113  &   -0.256  &    0.327  &   -0.353  &    0.468 \\
  NGC3109         &  0  &  0  &  -12.689  &   -0.077  &    0.076  &   -0.210  &    0.272  &   -0.286  &    0.411 \\
  NGC3198         &  1  &  0  &  -12.971  &   -0.090  &    0.066  &   -0.233  &    0.226  &   -0.283  &    0.319 \\
  NGC3521         &  1  &  1  &  -12.921  &   -0.120  &    0.107  &   -0.356  &    0.330  &   -0.530  &    0.552 \\
  NGC3726         &  1  &  0  &  -13.114  &   -0.172  &    0.129  &   -0.458  &    0.490  &   -0.593  &    0.724 \\
  NGC3741         &  1  &  1  &  -12.834  &   -0.057  &    0.089  &   -0.190  &    0.250  &   -0.255  &    0.367 \\
  NGC3769         &  1  &  1  &  -12.989  &   -0.138  &    0.154  &   -0.404  &    0.515  &   -0.568  &    0.889 \\
  NGC3877         &  0  &  0  &  -13.220  &   -0.261  &    0.308  &   -2.375  &    0.885  &   -5.404  &    1.107 \\
  NGC3893         &  1  &  1  &  -12.749  &   -0.131  &    0.156  &   -0.395  &    0.480  &   -0.509  &    0.796 \\
  NGC3917         &  1  &  0  &  -13.157  &   -0.187  &    0.148  &   -0.516  &    0.539  &   -0.727  &    0.820 \\
\hline
\hline
\end{tabular}
\end{center}
\end{table*}

\begin{table*}
\contcaption{}
\begin{center}
\begin{tabular}{l|ccccccccc}
\hline
\hline
  Galaxy          &$P_5$&$P_3$&$\log_{10} a_0 $ & $1\sigma_-$ & $1\sigma_+$ & $3\sigma_-$ & $3\sigma_+$ & $5\sigma_-$ & $5\sigma_+$\\
  \hline
  NGC3949         &  0  &  0  &  -12.893  &   -0.326  &    0.224  &   -1.523  &    0.818  &   -5.802  &    1.024 \\
  NGC3953         &  0  &  0  &  -13.665  &   -0.916  &    0.379  &   -5.107  &    0.698  &   -8.624  &    0.853 \\
  NGC3972         &  1  &  1  &  -12.803  &   -0.174  &    0.203  &   -0.511  &    0.667  &   -0.815  &    0.919 \\
  NGC3992         &  1  &  1  &  -13.275  &   -0.104  &    0.100  &   -0.301  &    0.313  &   -0.501  &    0.410 \\
  NGC4010         &  1  &  1  &  -12.805  &   -0.181  &    0.184  &   -0.525  &    0.619  &   -0.648  &    0.933 \\
  NGC4013         &  1  &  1  &  -12.884  &   -0.067  &    0.107  &   -0.210  &    0.321  &   -0.258  &    0.403 \\
  NGC4051         &  0  &  0  &  -13.586  &   -0.930  &    0.548  &   -5.343  &    1.009  &   -8.063  &    1.205 \\
  NGC4068         &  0  &  0  &  -13.566  &   -0.356  &    0.243  &   -1.146  &    0.954  &   -1.956  &    1.475 \\
  NGC4085         &  1  &  0  &  -12.788  &   -0.230  &    0.239  &   -0.713  &    0.795  &   -1.097  &    1.142 \\
  NGC4088         &  1  &  1  &  -13.149  &   -0.140  &    0.161  &   -0.409  &    0.498  &   -0.571  &    0.698 \\
  NGC4100         &  1  &  1  &  -13.206  &   -0.082  &    0.107  &   -0.265  &    0.310  &   -0.331  &    0.476 \\
  NGC4138         &  1  &  1  &  -13.024  &   -0.231  &    0.183  &   -0.693  &    0.639  &   -3.028  &    1.120 \\
  NGC4157         &  1  &  1  &  -12.970  &   -0.129  &    0.119  &   -0.356  &    0.410  &   -0.479  &    0.635 \\
  NGC4183         &  1  &  1  &  -13.437  &   -0.124  &    0.158  &   -0.388  &    0.473  &   -0.586  &    0.777 \\
  NGC4214         &  0  &  0  &  -12.321  &   -0.623  &    0.612  &   -1.141  &    2.268  &   -1.495  &    2.546 \\
  NGC4217         &  0  &  0  &  -12.524  &   -0.133  &    0.152  &   -0.411  &    0.531  &   -0.539  &    0.917 \\
  NGC4389         &  0  &  0  &  -13.362  &   -0.535  &    0.500  &   -4.423  &    2.018  &   -8.546  &    2.272 \\
  NGC4559         &  1  &  1  &  -12.948  &   -0.310  &    0.233  &   -0.639  &    1.072  &   -0.793  &    1.142 \\
  NGC5005         &  0  &  0  &  -12.797  &   -0.207  &    0.230  &   -0.896  &    0.622  &   -1.794  &    0.806 \\
  NGC5033         &  0  &  0  &  -13.178  &   -0.091  &    0.093  &   -0.299  &    0.275  &   -0.404  &    0.358 \\
  NGC5055         &  0  &  0  &  -12.868  &   -0.079  &    0.075  &   -0.281  &    0.182  &   -0.334  &    0.196 \\
  NGC5371         &  0  &  0  &  -13.852  &   -0.070  &    0.070  &   -0.211  &    0.206  &   -0.304  &    0.270 \\
  NGC5585         &  0  &  0  &  -12.534  &   -0.269  &    0.379  &   -0.767  &    1.088  &   -0.889  &    1.280 \\
  NGC5907         &  0  &  0  &  -13.390  &   -0.034  &    0.045  &   -0.102  &    0.130  &   -0.173  &    0.195 \\
  NGC5985         &  0  &  0  &  -13.822  &   -0.077  &    0.071  &   -0.223  &    0.221  &   -0.313  &    0.326 \\
  NGC6015         &  0  &  0  &  -13.542  &   -0.087  &    0.091  &   -0.256  &    0.276  &   -0.351  &    0.395 \\
  NGC6195         &  1  &  1  &  -13.115  &   -0.100  &    0.117  &   -0.305  &    0.339  &   -0.480  &    0.500 \\
  NGC6503         &  1  &  0  &  -12.825  &   -0.046  &    0.043  &   -0.126  &    0.145  &   -0.193  &    0.201 \\
  NGC6674         &  0  &  0  &  -13.527  &   -0.069  &    0.090  &   -0.215  &    0.258  &   -0.280  &    0.342 \\
  NGC6789         &  1  &  1  &  -12.319  &   -0.199  &    0.367  &   -0.733  &    1.125  &   -1.042  &    1.775 \\
  NGC6946         &  1  &  1  &  -13.061  &   -0.072  &    0.126  &   -0.284  &    0.306  &   -0.338  &    0.413 \\
  NGC7331         &  1  &  1  &  -12.781  &   -0.066  &    0.062  &   -0.183  &    0.212  &   -0.229  &    0.280 \\
  NGC7793         &  1  &  1  &  -13.295  &   -0.173  &    0.242  &   -0.521  &    0.815  &   -0.656  &    1.117 \\
  NGC7814         &  1  &  1  &  -12.574  &   -0.051  &    0.061  &   -0.162  &    0.176  &   -0.221  &    0.249 \\
  PGC51017        &  0  &  0  &  -15.216  &   -1.085  &    0.470  &   -4.461  &    0.965  &   -6.882  &    1.214 \\
  UGC00128        &  0  &  0  &  -14.072  &   -0.077  &    0.064  &   -0.206  &    0.219  &   -0.290  &    0.302 \\
  UGC00191        &  1  &  0  &  -13.893  &   -0.196  &    0.224  &   -0.635  &    0.762  &   -0.846  &    1.033 \\
  UGC00634        &  1  &  1  &  -13.008  &   -0.334  &    0.456  &   -0.891  &    1.328  &   -1.202  &    1.730 \\
  UGC00731        &  0  &  0  &  -12.713  &   -0.311  &    0.414  &   -0.725  &    1.446  &   -0.961  &    1.637 \\
  UGC00891        &  0  &  0  &  -12.596  &   -0.281  &    0.360  &   -0.786  &    1.046  &   -0.878  &    1.264 \\
  UGC01230        &  0  &  0  &  -15.361  &   -1.078  &    0.620  &   -4.471  &    1.723  &   -6.543  &    3.843 \\
  UGC01281        &  1  &  1  &  -12.995  &   -0.068  &    0.089  &   -0.232  &    0.236  &   -0.317  &    0.388 \\
  UGC02023        &  0  &  0  &  -13.451  &   -1.422  &    1.065  &   -5.055  &    3.715  &   -9.058  &    4.461 \\
  UGC02259        &  0  &  0  &  -13.623  &   -0.290  &    0.251  &   -0.810  &    0.954  &   -1.063  &    1.199 \\
  UGC02455        &  0  &  0  &  -13.292  &   -0.463  &    0.337  &   -4.016  &    1.699  &   -6.695  &    2.232 \\
  UGC02487        &  0  &  0  &  -13.467  &   -0.046  &    0.084  &   -0.182  &    0.214  &   -0.271  &    0.282 \\
  UGC02885        &  1  &  1  &  -13.002  &   -0.076  &    0.110  &   -0.251  &    0.318  &   -0.389  &    0.475 \\
  UGC02916        &  0  &  0  &  -13.616  &   -0.203  &    0.263  &   -0.829  &    0.803  &   -1.081  &    0.884 \\
  UGC02953        &  0  &  0  &  -12.976  &   -0.057  &    0.054  &   -0.151  &    0.144  &   -0.175  &    0.168 \\
  UGC03205        &  0  &  0  &  -13.227  &   -0.080  &    0.060  &   -0.240  &    0.193  &   -0.301  &    0.246 \\
  UGC03546        &  1  &  1  &  -12.919  &   -0.134  &    0.093  &   -0.378  &    0.332  &   -0.447  &    0.504 \\
  UGC03580        &  0  &  0  &  -12.360  &   -0.326  &    0.216  &   -0.640  &    0.722  &   -0.703  &    0.859 \\
  UGC04278        &  1  &  0  &  -12.650  &   -0.277  &    0.400  &   -0.606  &    1.810  &   -0.832  &    1.853 \\
  UGC04305        &  0  &  0  &  -15.526  &   -0.973  &    0.731  &   -4.201  &    1.836  &   -7.082  &    2.361 \\
  UGC04325        &  0  &  0  &  -13.041  &   -0.295  &    0.362  &   -2.365  &    1.335  &   -3.097  &    1.391 \\
  UGC04483        &  1  &  1  &  -13.411  &   -0.117  &    0.156  &   -0.398  &    0.439  &   -0.583  &    0.607 \\
  UGC04499        &  1  &  1  &  -13.303  &   -0.251  &    0.367  &   -0.767  &    1.270  &   -1.073  &    1.543 \\
  UGC05005        &  1  &  1  &  -13.166  &   -0.496  &    0.397  &   -1.148  &    1.735  &   -1.641  &    2.361 \\
  UGC05253        &  0  &  0  &  -12.753  &   -0.075  &    0.071  &   -0.224  &    0.203  &   -0.285  &    0.273 \\
  UGC05414        &  1  &  1  &  -13.189  &   -0.337  &    0.309  &   -0.844  &    1.227  &   -1.338  &    1.509 \\
  UGC05716        &  0  &  0  &  -13.871  &   -0.191  &    0.176  &   -0.498  &    0.651  &   -0.730  &    0.823 \\
  UGC05721        &  1  &  1  &  -12.669  &   -0.212  &    0.262  &   -0.541  &    0.851  &   -0.727  &    1.047 \\
  UGC05750        &  1  &  1  &  -13.314  &   -0.301  &    0.269  &   -0.789  &    1.062  &   -1.187  &    1.708 \\
  UGC05764        &  0  &  0  &  -12.945  &   -0.343  &    0.272  &   -0.760  &    0.932  &   -0.912  &    1.114 \\
\hline
\hline
\end{tabular}
\end{center}
\end{table*}

\begin{table*}
\contcaption{}
\begin{center}
\begin{tabular}{l|ccccccccc}
\hline
\hline
  Galaxy          &$P_5$&$P_3$&$\log_{10} a_0 $ & $1\sigma_-$ & $1\sigma_+$ & $3\sigma_-$ & $3\sigma_+$ & $5\sigma_-$ & $5\sigma_+$\\
  \hline
  UGC05829        &  1  &  1  &  -13.084  &   -0.490  &    0.710  &   -1.400  &    2.278  &   -2.139  &    3.078 \\
  UGC05918        &  1  &  1  &  -13.150  &   -0.270  &    0.429  &   -0.842  &    1.481  &   -1.234  &    1.775 \\
  UGC05986        &  1  &  0  &  -12.549  &   -0.187  &    0.343  &   -0.495  &    1.267  &   -0.702  &    1.899 \\
  UGC05999        &  0  &  0  &  -13.264  &   -0.514  &    1.090  &   -1.752  &    2.995  &   -2.707  &    3.309 \\
  UGC06399        &  1  &  1  &  -12.859  &   -0.154  &    0.164  &   -0.449  &    0.545  &   -0.590  &    0.849 \\
  UGC06446        &  1  &  1  &  -13.054  &   -0.280  &    0.270  &   -0.732  &    1.182  &   -1.033  &    1.369 \\
  UGC06614        &  1  &  1  &  -12.537  &   -0.376  &    0.422  &   -0.941  &    1.569  &   -1.098  &    1.870 \\
  UGC06628        &  0  &  0  &  -14.446  &   -1.440  &    1.302  &   -4.975  &    4.232  &   -6.976  &    5.263 \\
  UGC06667        &  1  &  0  &  -12.523  &   -0.135  &    0.137  &   -0.357  &    0.491  &   -0.463  &    0.724 \\
  UGC06786        &  1  &  1  &  -12.094  &   -0.163  &    0.107  &   -0.385  &    0.335  &   -0.449  &    0.432 \\
  UGC06787        &  0  &  0  &  -13.461  &   -0.051  &    0.054  &   -0.137  &    0.261  &   -0.172  &    0.320 \\
  UGC06818        &  1  &  0  &  -12.939  &   -0.182  &    0.392  &   -0.603  &    0.832  &   -0.817  &    1.176 \\
  UGC06917        &  1  &  1  &  -12.850  &   -0.196  &    0.130  &   -0.511  &    0.506  &   -0.784  &    0.753 \\
  UGC06923        &  1  &  1  &  -13.060  &   -0.172  &    0.211  &   -0.562  &    0.645  &   -0.857  &    0.966 \\
  UGC06930        &  1  &  1  &  -13.393  &   -0.295  &    0.387  &   -1.037  &    1.194  &   -1.468  &    1.794 \\
  UGC06973        &  0  &  0  &  -12.023  &   -0.146  &    0.144  &   -0.395  &    0.519  &   -0.505  &    0.789 \\
  UGC06983        &  1  &  1  &  -12.982  &   -0.115  &    0.173  &   -0.390  &    0.511  &   -0.488  &    1.086 \\
  UGC07089        &  1  &  1  &  -13.306  &   -0.199  &    0.179  &   -0.563  &    0.642  &   -0.819  &    1.001 \\
  UGC07125        &  1  &  1  &  -13.994  &   -0.239  &    0.326  &   -0.707  &    1.130  &   -1.206  &    1.474 \\
  UGC07151        &  1  &  1  &  -13.162  &   -0.109  &    0.076  &   -0.300  &    0.261  &   -0.432  &    0.353 \\
  UGC07232        &  0  &  0  &  -12.844  &   -0.205  &    0.188  &   -0.616  &    0.622  &   -1.084  &    0.753 \\
  UGC07261        &  1  &  1  &  -13.324  &   -0.550  &    0.724  &   -1.986  &    2.281  &   -2.765  &    2.878 \\
  UGC07323        &  1  &  1  &  -13.170  &   -0.337  &    0.399  &   -1.111  &    1.358  &   -2.012  &    1.646 \\
  UGC07399        &  1  &  1  &  -12.322  &   -0.254  &    0.257  &   -0.567  &    1.019  &   -0.816  &    1.174 \\
  UGC07524        &  1  &  1  &  -13.248  &   -0.101  &    0.149  &   -0.334  &    0.424  &   -0.445  &    0.640 \\
  UGC07559        &  0  &  0  &  -13.615  &   -0.154  &    0.155  &   -0.494  &    0.460  &   -0.771  &    0.582 \\
  UGC07577        &  0  &  0  &  -14.543  &   -0.554  &    0.349  &   -4.359  &    0.746  &   -7.555  &    0.934 \\
  UGC07603        &  1  &  1  &  -12.579  &   -0.248  &    0.321  &   -0.567  &    1.118  &   -0.798  &    1.248 \\
  UGC07608        &  0  &  0  &  -12.930  &   -0.555  &    0.811  &   -1.611  &    2.950  &   -3.897  &    3.455 \\
  UGC07690        &  1  &  1  &  -13.545  &   -0.353  &    0.387  &   -1.243  &    1.402  &   -3.547  &    2.267 \\
  UGC07866        &  0  &  0  &  -13.450  &   -0.269  &    0.227  &   -0.835  &    0.778  &   -1.358  &    1.305 \\
  UGC08286        &  1  &  0  &  -12.865  &   -0.049  &    0.043  &   -0.139  &    0.138  &   -0.189  &    0.181 \\
  UGC08490        &  1  &  1  &  -12.974  &   -0.121  &    0.122  &   -0.335  &    0.392  &   -0.472  &    0.576 \\
  UGC08550        &  1  &  1  &  -13.104  &   -0.171  &    0.334  &   -0.509  &    1.043  &   -0.719  &    1.225 \\
  UGC08699        &  1  &  1  &  -12.986  &   -0.100  &    0.162  &   -0.339  &    0.434  &   -0.442  &    0.586 \\
  UGC08837        &  0  &  0  &  -13.517  &   -0.170  &    0.118  &   -0.487  &    0.364  &   -0.764  &    0.489 \\
  UGC09037        &  1  &  1  &  -12.870  &   -0.148  &    0.122  &   -0.394  &    0.419  &   -0.512  &    0.679 \\
  UGC09133        &  0  &  0  &  -13.289  &   -0.045  &    0.050  &   -0.201  &    0.141  &   -0.212  &    0.188 \\
  UGC09992        &  0  &  0  &  -13.884  &   -1.364  &    0.951  &   -5.280  &    2.972  &   -7.462  &    4.018 \\
  UGC10310        &  1  &  1  &  -13.353  &   -0.388  &    0.607  &   -1.896  &    1.914  &   -2.739  &    2.241 \\
  UGC11455        &  0  &  0  &  -12.855  &   -0.076  &    0.113  &   -0.246  &    0.334  &   -0.361  &    0.441 \\
  UGC11557        &  0  &  0  &  -13.319  &   -0.678  &    0.791  &   -3.596  &    2.781  &   -6.638  &    2.982 \\
  UGC11820        &  1  &  1  &  -13.164  &   -0.393  &    0.304  &   -0.877  &    1.064  &   -1.155  &    1.202 \\
  UGC11914        &  1  &  1  &  -12.183  &   -0.140  &    0.096  &   -0.405  &    0.311  &   -0.554  &    0.385 \\
  UGC12506        &  1  &  1  &  -13.345  &   -0.106  &    0.128  &   -0.342  &    0.360  &   -0.514  &    0.535 \\
  UGC12632        &  1  &  1  &  -13.466  &   -0.280  &    0.332  &   -0.795  &    1.161  &   -1.228  &    1.385 \\
  UGC12732        &  1  &  1  &  -13.441  &   -0.370  &    0.374  &   -0.984  &    1.201  &   -1.664  &    1.476 \\
  UGCA281         &  0  &  0  &  -13.149  &   -0.130  &    0.115  &   -0.379  &    0.368  &   -0.563  &    0.541 \\
  UGCA442         &  1  &  0  &  -12.839  &   -0.105  &    0.198  &   -0.302  &    0.617  &   -0.390  &    0.793 \\
  UGCA444         &  1  &  1  &  -13.272  &   -0.062  &    0.091  &   -0.215  &    0.257  &   -0.303  &    0.402 \\
\hline
\hline
\end{tabular}
\end{center}
\end{table*}


\bsp	
\label{lastpage}
\end{document}